\documentclass[eqsecnum,showpacs,showkeys,nofootinbib,aps]{revtex4}
\usepackage{epsfig}

\renewcommand{\theequation}{\arabic{equation}}
\def\be{\begin{equation}}
\def\ee{\end{equation}}
\def\bea{\begin{eqnarray}}
\def\eea{\end{eqnarray}}

\begin{document}

\title{Tidal effects in Schwarzschild black hole in holographic massive gravity }
\author{Soon-Tae Hong}
\email{galaxy.mass@gmail.com}
\affiliation{Center for Quantum Spacetime and
\\ Department of Physics, Sogang University, Seoul 04107, Korea}
\author{Yong-Wan Kim}
\email{ywkim65@gmail.com}
 \affiliation{Department of Physics and
 \\ Research Institute of Physics and Chemistry, Jeonbuk National University, Jeonju 54896, Korea}
\author{Young-Jai Park}
\affiliation{Center for Quantum Spacetime and
\\ Department of Physics, Sogang University, Seoul 04107, Korea }
\date{\today}

\begin{abstract}
We investigate tidal effects produced in the spacetime of
Schwarzschild black hole in holographic massive gravity, which has
two additional mass parameters due to massive gravitons. As a
result, we have obtained that massive gravitons affect the angular
component of the tidal force, while the radial component has the
same form with the one in massless gravity. On the other hand, by
solving the geodesic deviation equations, we have found that
radial components of two nearby geodesics keep tightening while
falling into the black hole and after passing the event horizons
get abruptly infinitely stretched due to massive gravitons.
However, angular components of two nearby geodesics get stretched
firstly, reach a peak and then get compressed while falling into
the black hole. Moreover, we have also shown that the angular
components are more easily deformed near the departure position as
the mass of a black hole is smaller for a fixed graviton mass.
\end{abstract}


\maketitle

\section{introduction}
\setcounter{equation}{0}
\renewcommand{\theequation}{\arabic{section}.\arabic{equation}}

Einstein's theory of general relativity (GR) has been tested
successfully to date as the description of the force of gravity.
Despite all the successes of GR, some puzzles
not only on cosmological scales but also on quantum scales have
pushed forward to search for alternatives. Various modifications
are obtained by adding extra scalar, vector, or tensor fields to
their gravitational sector, and additional scalar curvature
invariants to the action of GR \cite{Clifton:2011jh}.

In particular, introducing extra tensor fields as a background can
make Einstein's massless spin-2 graviton to be massive
\cite{Hinterbichler:2011tt,deRham:2014zqa}. For example, a massive
spin-2 theory can be considered as a theory with a dynamical
fluctuation on a nondynamical background. Taking the background
tensor to be the Minkowskian one, one can have a massive graviton
by adding the Pauli-Fierz mass term to the GR action, resulting in
the Pauli-Fierz action \cite{Fierz:1939ix}. However, it was later
known that the massive gravity suffered from the Boulware-Deser
ghost problem \cite{Boulware:1973my} and the van Dam, Veltman and
Zakharov (vDVZ) discontinuity \cite{vanDam:1970vg,Zakharov:1970cc}
in the massless graviton limit.

A decade ago, de Rham, Gabadadze and Trolley (dRGT)
\cite{deRham:2010ik,deRham:2010kj} obtained a ghost free massive
gravity, which has nonlinearly interacting mass terms constructed
from the metric coupled with a symmetric background tensor, called
the reference metric. In the dRGT massive gravity, the
nondynamical background tensor is also set to be the Minkowskian
one. In order to preserve diffeomorphism invariance, Hassan et al.
\cite{Hassan:2011hr,Hassan:2011tf} developed the ghost free
massive gravity with a general reference metric. On the other
hand, Vegh \cite{Vegh:2013sk} introduced a nonlinear massive
gravity with a special singular reference metric as a background
tensor, which is used to study momentum dissipation for describing
the electric and heat conductivity for normal conductors. The
nondynamical background tensor is chosen to keep the
diffeomorphism symmetry for coordinates ($t,r$) intact, but breaks
it in angular directions. Due to a broken momentum conservation,
graviton acquires the mass that leads to momentum dissipation in
the dual holographic theory. Thus, massive gravity theories with
broken diffeomorphism invariance in the bulk provide a holographic
model for theories with broken spatial translational symmetry at
the boundary \cite{Davison:2013jba,Blake:2013bqa,Blake:2013owa}.
Since then, this has been extensively exploited to investigate
many black hole models
\cite{Cai:2014znn,Adams:2014vza,Hendi:2015pda,Hu:2016hpm,Zou:2016sab,
Hendi:2017fxp,Tannukij:2017jtn,Hendi:2017bys,Hendi:2018xuy,Chabab:2019mlu,Hong:2018spz,Hong:2019zsi}
as well as holographic condensed matter physics
\cite{Andrade:2013gsa,Amoretti:2014zha,Baggioli:2014roa,Zhou:2015dha,Hartnoll:2016apf,
Alberte:2017oqx,Ammon:2019wci}.
In addition, it was also studied what effects of the holographic
massive gravity were on the structure of physical neutron stars
\cite{Hendi:2017ibm}.

On the other hand, it is well-known that a body in free fall
toward the center of another body gets stretched in the radial
direction and compressed in the angular one. The stretching and
compression arise from a tidal effect of gravity, which is given
by a difference in the strength of gravity between two points
\cite{MTW:1973,DInverno:1992,Carroll:2004,Hobson:2006}. Tidal
phenomena are common in the universe from our solar system to
stars in binary systems, to galaxies, to cluster of galaxies, and
even to gravitational waves \cite{Goswami:2019fyk}. In particular,
Wheeler \cite{Wheeler:1971} proposed that a star in the ergosphere
of the Kerr black hole can be broken up due to tidal interaction
and emit subsequently a jet composed of the debris as a mechanism
for the jets production. Since then, the investigation of tidal
effects in astrophysical context has been devoted to tidal
disruption of stars deeply plunging into black holes
\cite{Hills:1975,Carter:1982,Rees:1988,komossa:2015,
Auchettl:2017,Rossi:2020rvv}. Many other studies in theoretical
context have also been preformed for the extensive description of
tidal effects in various black holes
\cite{Mahajan:1981,AbdelMegied:2004ni,Crispino:2016pnv,Gad2010,Shahzad:2017vwi,
Chan:1995fc,Nandi:2000gt,Cardoso:2012zn,Harko:2012ve,Uniyal:2014oaa,
Sharif:2018a,Sharif:2018gzj,Junior:2020yxg,Junior:2020par}. Very
recently, making use of a tidal acceleration of the separation
between two arms of gravitational wave detectors such as LIGO
\cite{Abbott:2016blz,Abbott:2017vtc}, the authors  have proposed a
new method of a direct measurement of gravitons
\cite{Parikh:2020nrd,Parikh:2020kfh,Parikh:2020fhy}.


However, studies on tidal effects in astrophysical context so far
have been mainly devoted to either the nonrotating Schwarzschild
or the rotating Kerr black holes in massless gravity. Moreover, in
theoretical context, even though tidal effects have also been
studied in various black holes, there are few works on possible
deformation of geodesic deviation vectors when the metric is
changed according to massive graviton.

Motivated by this, we will study tidal effects in the
Schwarzschild black hole in Vegh's holographically massive
gravity, which has two additional mass parameters due to massive
gravitons, comparing them with the Schwarzschild black hole in
massless gravity. In Sec. II and III, we investigate features of
the geodesic equations and tidal forces for the Schwarzschild
black hole in holographic massive gravity. In Sec. IV, we find
solutions of the geodesic deviation equations for radially falling
bodies toward the Schwarzschild black hole in holographic massive
gravity and analyze the results having massive gravitons comparing
with the ones in massless gravity. Conclusions are drawn in Sec.
V.

\section{Geodesics in Schwarzschild black hole in holographic massive gravity}
\setcounter{equation}{0}
\renewcommand{\theequation}{\arabic{section}.\arabic{equation}}


In this section, we will newly study the geodesic equations of
Schwarzschild black hole in holographic massive gravity, which is
described by the action
 \be\label{mSchads}
 S=\frac{1}{16\pi G}\int d^4x\sqrt{-g}\left[{\cal R}
   +{\tilde m}^2\sum_{i=1}^{4}a_i{\cal
   U}_{i}(g_{\mu\nu},f_{\mu\nu})\right],
 \ee
where ${\cal R}$ is the scalar curvature, $\tilde{m}$ is the
graviton mass\footnote{In particular, we will call it massless
when $\tilde{m}$ is zero in this work.}, $a_i$ are constants and
${\cal U}_i$ are symmetric polynomials of the eigenvalue of the
matrix ${\cal K}^\mu_\nu\equiv\sqrt{g^{\mu\alpha}f_{\alpha\nu}}$
given by
 \bea
 {\cal U}_1 &=& [{\cal K}],
 ~~~{\cal U}_2 = [{\cal K}]^2-[{\cal K}^2],
 ~~~{\cal U}_3 = [{\cal K}]^3-3[{\cal K}][{\cal K}^2]+2[{\cal K}^3]\nonumber\\
 {\cal U}_4 &=& [{\cal K}]^4-6[{\cal K}^2][{\cal K}]^2+8[{\cal K}^3][{\cal K}]+3[{\cal K}^2]^2-6[{\cal K}^4].
 \eea
The square root in ${\cal K}$ means
$(\sqrt{A})^\mu_\alpha(\sqrt{A})^\alpha_\nu=A^\mu_\nu$ and $[{\cal
K}]$ denotes the trace ${\cal K}^\mu_\mu$. Finally, $f_{\mu\nu}$,
called the reference metric, is a non-dynamical, fixed symmetric
tensor introduced to construct nontrivial interaction terms in
holographic massive gravity. Then, with a gauge-fixed ansatz for
the reference metric as
 \be\label{fidmetric}
 f_{\mu\nu}={\rm diag}(0,0,a^2_0,a^2_0\sin^2\theta),
 \ee
where $a_0$ is a positive constant
\cite{Vegh:2013sk,Blake:2013bqa,Amoretti:2014zha,Zhou:2015dha,Cai:2014znn,Adams:2014vza,Hong:2018spz,Hong:2019zsi},
one can find the spherically symmetric black hole solution as
 \bea\label{metric-mSch}
 ds^2= g_{\mu\nu}dx^\mu dx^\nu = -f(r)dt^2+f^{-1}(r)dr^2+r^2(d\theta^2+\sin^2\theta d\phi^2)
 \eea
with
 \be\label{lapsewol}
 f(r)=1-\frac{2m}{r}+2Rr+{\cal C}.
 \ee
Here, we have newly defined $R=a_0a_1\tilde{m}^2/4$ and ${\cal
C}=a^2_0a_2\tilde{m}^2$ without loss of generality.

\begin{figure*}[t!]
   \centering
   \includegraphics{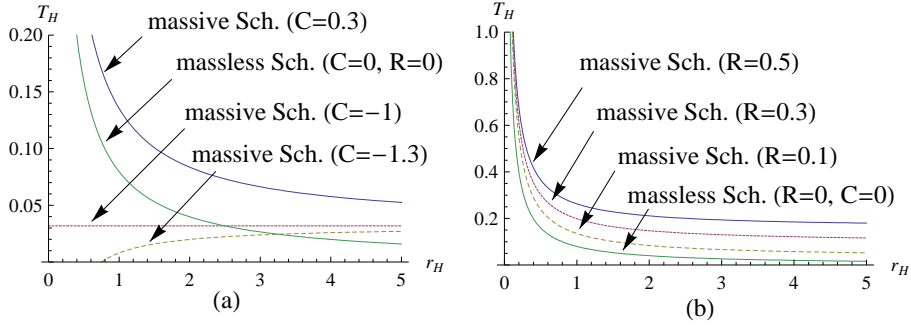}
\caption{Hawking temperature for the Schwarzschild black hole in
holographic massive gravity: (a) varying $\cal C$ with $R=0.1$ and
(b) varying $R$ with $C=0.3$, comparing with ones in massless
gravity.}
 \label{fig1}
\end{figure*}

In order to see the difference between the Schwarzschild black
holes in holographic massive gravity and in massless one, we have
drawn the mass parameter dependent Hawking temperature in Fig.
\ref{fig1}
 \be\label{HTmSch}
 T_H=\frac{1+{\cal C}}{4\pi r_H}+\frac{R}{\pi},
 \ee
which shows the role of two additional mass terms $R$ and ${\cal
C}$ clearly. As seen from Fig. \ref{fig1}, the Hawking temperature
in the holographic massive gravity differently behaves according
to the values of ${\cal C}$ and $R$. Note that first $R$ gives a
constant contribution to the Hawking temperature. The Hawking
temperature is mostly proportional to $1/r_H$. When ${\cal C}>-1$,
it decreases as $r_H$ increases. When ${\cal C}=-1$, it is just a
constant given by $R$. However, when ${\cal C}<-1$, it is
negatively proportional to $1/r_H$. Note that the Hawking
temperature is flipped by the sign of $\cal C$. In this paper, we
assume that the mass parameters are positive without loss of
generality.

Now, the geodesic equations of
 \be\label{gdeq}
 \frac{d^2x^\mu}{d\tau^2}+\Gamma^\mu_{\nu\rho}\frac{dx^\nu}{d\tau}\frac{dx^\rho}{d\tau}=0
 \ee
can be obtained from the metric where $x^\mu=(t,r,\theta,\phi)$.
Here, the independent non-vanishing components of the Christoffel
symbols are
 \bea\label{csymbols}
 \Gamma^0_{01}&=&-\Gamma^1_{11}=\frac{f'(r)}{2f(r)},
   ~~~\Gamma^1_{00}=\frac{1}{2}f'(r)f(r),
   ~~~~~\Gamma^1_{22}=-r f(r),
   ~~~\Gamma^1_{33}=-r f(r)\sin^2\theta, \nonumber\\
   \Gamma^2_{12}&=&\Gamma^3_{13}=\frac{1}{r},
   ~~~~~~~~~~~\Gamma^2_{33}=-\sin\theta\cos\theta,
   ~~~~\Gamma^3_{23}=\cot\theta.
 \eea
From these equations, one can explicitly obtain the geodesic
equations as
 \bea
 &&\frac{dv^0}{d\tau}+\frac{2(m+Rr^2)}{r^2-2mr+2Rr^3+{\cal C}r^2}v^0v^1=0, \\
 &&\frac{dv^1}{d\tau}+\frac{(m+Rr^2)}{r^4}(r^2-2mr+2Rr^3+{\cal C}r^2)(v^0)^2
      -\frac{m+Rr^2}{r^2-2mr+2Rr^3+{\cal C}r^2}(v^1)^2
      \nonumber\\
      &&-\frac{r^2-2mr+2Rr^3+{\cal C}r^2}{r}[(v^2)^2+\sin^2\theta(v^3)^2]=0,     \\
 &&\frac{dv^2}{d\tau}+\frac{2}{r}v^1v^2-\sin\theta\cos\theta(v^3)^2=0, \\
 &&\frac{dv^3}{d\tau}+\frac{2}{r}v^1v^3+\cot\theta v^2 v^3=0
 \eea
in terms of the four velocity vector $v^\mu=dx^\mu/d\tau$. Without
loss of generality, one can consider the geodesics on the
equatorial plane $\theta=\pi/2$ for all $\tau$. Then, one has
$v^2=\dot\theta=0$ and the geodesic equations are finally reduced
to
 \bea
 &&\frac{dv^0}{d\tau}+\frac{2(m+Rr^2)}{r^2-2mr+2Rr^3+{\cal C}r^2}v^0v^1=0, \label{ge0m}\\
 &&\frac{dv^1}{d\tau}+\frac{(m+Rr^2)}{r^4}(r^2-2mr+2Rr^3+{\cal C}r^2)(v^0)^2
       -\frac{m+Rr^2}{r^2-2mr+2Rr^3+{\cal C}r^2}(v^1)^2 \nonumber\\
      &&-\frac{r^2-2mr+2Rr^3+{\cal C}r^2}{r}(v^3)^2=0,     \\
 &&\frac{dv^3}{d\tau}+\frac{2}{r}v^1v^3=0\label{ge2m}.
 \eea
By making use of $v^1=\dot{r}$, one can integrate Eqs.
(\ref{ge0m}) and (\ref{ge2m}) as
 \bea
 v^0=\frac{c_1 r^2}{r^2-2mr+2Rr^3+{\cal C}r^2}, ~~~
 v^3=\frac{c_2}{r^2}, \label{v0mand2m}
 \eea
respectively, where $c_1$ and $c_2$ are integration constants. It
seems appropriate to comment that for the Killing vectors
$\xi^\mu=(1,0,0,0)$ and $\psi^\mu=(0,0,0,1)$, two conserved
quantities are given by
 \bea
 E = -g_{\mu\nu}\xi^\mu v^\nu =f(r) v^0,
 ~~~L = g_{\mu\nu}\psi^\mu v^\nu =r^2 v^3.\label{K1}
 \eea
Comparing these relations with Eqs. (\ref{v0mand2m}), we can fix
the integration constants as $c_1=E$, $c_2=L$ in terms of the
conserved quantities of $E$ and $L$.

Finally, by letting $ds^2=-kd\tau^2$ in Eq. (\ref{metric-mSch})
and using Eqs. (\ref{v0mand2m}), one can obtain
 \be
 v^1=\pm\left[E^2-\left(k+\frac{L^2}{r^2}\right)\left(1-\frac{2m}{r}+2Rr+{\cal C}\right)\right]^{1/2},
 \ee
where $-/+$ sign is for inward/outward motion as before. Moreover,
timelike (nulllike) geodesic is for $k=1~(0)$. Note that in the
massless limit of both $R=0$ and ${\cal C}=0$, one can easily
reproduce the previous geodesic results of the Schwarzschild black
hole in massless gravity
\cite{MTW:1973,DInverno:1992,Carroll:2004,Hobson:2006}.

\section{Tidal force in the Schwarzschild
black hole in holographic massive gravity}
\setcounter{equation}{0}
\renewcommand{\theequation}{\arabic{section}.\arabic{equation}}

Now, let us consider the tidal force acting in the Schwarzschild
black hole in holographic massive gravity. First of all, let us
define the geodesic deviation, or separation four-vectors
$\eta^\mu$ which denote the infinitesimal displacement between two
nearby particles in free fall. Then, the equations of the geodesic
deviation \cite{MTW:1973,DInverno:1992,Carroll:2004,Hobson:2006}
are given by
 \be\label{gd}
 \frac{D^2\eta^\mu}{D\tau^2}+R^\mu_{\nu\rho\sigma}v^\nu
 \eta^\rho v^\sigma = 0,
 \ee
where $R^\mu_{\nu\rho\sigma}$ is the Riemann curvature and $v^\mu$
is the unit tangent vector to the geodesic line.

In order to study the behavior of the separation vector in detail,
we consider the timelike geodesic equation with $L=0$ for
simplicity. We also introduce the tetrad basis describing a freely
falling frame given by
 \bea
 e^\mu_{\hat 0}&=&\left(\frac{E}{f(r)},-\sqrt{E^2-f(r)},0,0\right),~~
 e^\mu_{\hat 1}=\left(-\frac{\sqrt{E^2-f(r)}}{f(r)},E,0,0\right),\nonumber\\
 e^\mu_{\hat 2}&=&\left(0,0,\frac{1}{r},0\right),~~~~~~~~~~~~~~~~~~~~~~
 e^\mu_{\hat 3}=\left(0,0,0,\frac{1}{r\sin\theta}\right),
 \eea
satisfying the orthonormality relation of $e^\mu_{\hat\alpha}
e_{\mu\hat\beta}=\eta_{\hat\alpha\hat\beta}$ with
$\eta_{\hat\alpha\hat\beta}={\rm diag}(-1,1,1,1)$. The separation
vector can also be expanded as
$\eta^\mu=e^\mu_{\hat\alpha}\eta^{\hat\alpha}$ with a fixed
temporal component of $\eta^{\hat 0}=0$
\cite{DInverno:1992,Hobson:2006}.

In the tetrad basis, the Riemann tensor can be written as
 \be
 R^{\hat\alpha}_{\hat\beta\hat\gamma\hat\delta}
 =e^{\hat\alpha}_\mu e^\nu_{\hat\beta} e^\rho_{\hat\gamma}
 e^\sigma_{\hat\delta} R^\mu_{\nu\rho\sigma},
 \ee
so one can obtain the non-vanishing independent components of the
Riemann tensor  in holographic massive gravity as
 \bea
 R^{\hat 0}_{\hat 1 \hat 0 \hat 1}&=&-\frac{f''(r)}{2},
    ~~R^{\hat 0}_{\hat 2 \hat 0 \hat 2}=
      R^{\hat 0}_{\hat 3 \hat 0 \hat 3}=
      R^{\hat 1}_{\hat 2 \hat 1 \hat 2}=
      R^{\hat 1}_{\hat 3 \hat 1 \hat 3}=-\frac{f'(r)}{2r},
    ~~R^{\hat 2}_{\hat 3 \hat 2 \hat 3}=\frac{1-f(r)}{r^2}.
 \eea
Then, one can obtain the desired tidal forces in the radially
freely falling frame as
 \bea
 \frac{d^2\eta^{\hat 1}}{d\tau^2} &=& -\frac{f''(r)}{2}\eta^{\hat 1}=\frac{2m}{r^3}\eta^{\hat 1},\label{gdem1}\\
 \frac{d^2\eta^{\hat i}}{d\tau^2} &=& -\frac{f'(r)}{2r}\eta^{\hat i}=
                              -\left(\frac{m}{r^3}+\frac{R}{r}\right)\eta^{\hat i},\label{gdem2}
 \eea
where $i=2,~3$. Thus, one can find that comparing with the case of
the Schwarzschild black hole in massless gravity, the tidal effect
in the radial direction has exactly the same form
\cite{Mahajan:1981,AbdelMegied:2004ni,Crispino:2016pnv}. However,
we have newly obtained that the tidal effect in the angular
direction has additional term proportional to $R/r$ in the
Schwarzschild black hole in holographic massive gravity. Note that
the tidal forces are independent of the constant term ${\cal C}$
in $f(r)$ because these forces are obtained from $f'(r)$ and
$f''(r)$. Thus, the massive graviton effect seems to appear only
in the angular direction in the tidal force.

\section{Geodesic deviation equations of the Schwarzschild
black hole in holographic massive gravity}
\setcounter{equation}{0}
\renewcommand{\theequation}{\arabic{section}.\arabic{equation}}

In this section, we solve the geodesic deviation equations of
(\ref{gdem1}) and (\ref{gdem2}), and find the behavior of the
geodesic deviation vectors of test particles freely falling into
the Schwarzschild black hole in holographic massive gravity. Then,
we will compare this with the case of the Schwarzschild black hole
in massless gravity in order to explicitly show the massive
graviton effects on both the radial and angular directions.

For the geodesic deviation equations (\ref{gdem1}) and
(\ref{gdem2}) of the Schwarzschild black hole in holographic
massive gravity, the tidal forces can be rewritten in terms of
$r$-derivative as
 \bea\label{diffm1}
 [E^2-f(r)]\frac{d^2\eta^{\hat 1}}{dr^2}
  - \frac{f'(r)}{2}\frac{d\eta^{\hat 1}}{dr}+\frac{f''(r)}{2}\eta^{\hat
  1}&=&0,\\
 \label{diffm2}
 [E^2-f(r)]
 \frac{d^2\eta^{\hat i}}{dr^2}
  - \frac{f'(r)}{2}\frac{d\eta^{\hat i}}{dr}+\frac{f'(r)}{2r}\eta^{\hat i}&=&0.
 \eea
The solution of the radial component (\ref{diffm1}) is given by
 \be\label{rintegral}
 \eta^{\hat 1}(r)= d_1\sqrt{E^2-f(r)}
       +d_2\sqrt{E^2-f(r)}\int\frac{dr}{[E^2-f(r)]^{3/2}},
 \ee
and the angular component (\ref{diffm2}) is
 \be\label{aintegral}
 \eta^{\hat i}(r)=
    r\left(d_3+d_4\int\frac{dr}{r^2\sqrt{E^2-f(r)}}\right),
 \ee
where $d_1,\cdot\cdot\cdot, d_4$ are constants of integration.

At this stage, let us first recapitulate the integration with
$R=0$ and ${\cal C}=0$, which corresponds to the massless case
\cite{Mahajan:1981,AbdelMegied:2004ni,Crispino:2016pnv}. Then, the
geodesic deviation equations in Eqs. (\ref{diffm1}) and
(\ref{diffm2}) are reduced to
 \bea
 2m\left(\frac{1}{r}-\frac{1}{b}\right)\frac{d^2\eta^{\hat 1}}{dr^2}
  - \frac{m}{r^2}\frac{d\eta^{\hat 1}}{dr}-\frac{2m}{r^3}\eta^{\hat
  1}&=&0,\label{diff1}\\
  2m\left(\frac{1}{r}-\frac{1}{b}\right)\frac{d^2\eta^{\hat i}}{dr^2}
  - \frac{m}{r^2}\frac{d\eta^{\hat i}}{dr}+\frac{m}{r^3}\eta^{\hat
  i}&=&0.\label{diff2}
 \eea
Here, we are considering a body released from rest at $r=b$ so we
have $E=(1-2m/b)^{1/2}$. The solution of the radial component
(\ref{diff1}) is given by
 \be
 \eta^{\hat 1}(r)=\sqrt{2m\left(\frac{1}{r}-\frac{1}{b}\right)}
    \left(d_1+d_2\int\frac{dr}{[2m\left(\frac{1}{r}-\frac{1}{b}\right)]^{3/2}}\right),
 \ee
and the angular component (\ref{diff2}) is
 \be
 \eta^{\hat i}(r)=
    r\left(d_3+d_4\int\frac{dr}{r^2\sqrt{2m\left(\frac{1}{r}-\frac{1}{b}\right)}}\right),
 \ee
where $d_1,\cdot\cdot\cdot, d_4$ are constants of integration.
\begin{figure*}[t!]
   \centering
   \includegraphics{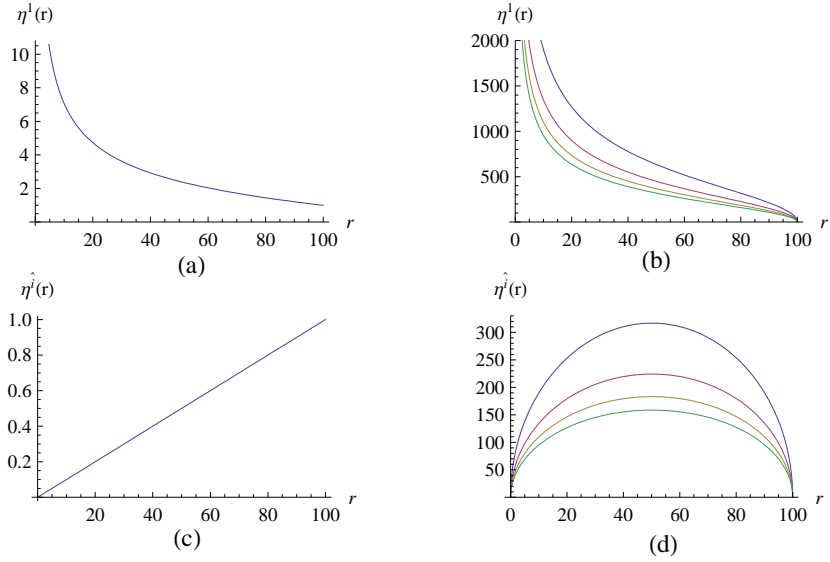}
\caption{For the Schwarzschild black hole in massless gravity,
radial solutions of the geodesic deviation equation: (a) for
$\eta^{\hat 1}(b)=1$, $\frac{d\eta^{\hat 1}(b)}{d\tau}$=0 with
$m=10$, (b) for $\eta^{\hat 1}(b)=1$, $\frac{d\eta^{\hat
1}(b)}{d\tau}$=1, $m=5,~10,~15,~20$ from bottom to top, and
angular solutions of the geodesic deviation equation, (c) for
$\eta^{\hat i}(b)=1$, $\frac{d\eta^{\hat i}(b)}{d\tau}$=0 with
$m=10$, and (d) for $\eta^{\hat i}(b)=1$, $\frac{d\eta^{\hat
i}(b)}{d\tau}$=1, $m=5,~10,~15,~20$ from top to bottom.}
 \label{fig2}
\end{figure*}
Since we are considering a body falling from rest at $r=b$ , we
find the constants of integration as
 \bea
 d_1=\frac{b^2}{m}\frac{d\eta^{\hat 1}(b)}{d\tau},~~d_2=\frac{m}{b^2}\eta^{\hat 1}(b),
 ~~d_3=\frac{1}{b}\eta^{\hat i}(b),~~d_4=-b\frac{d\eta^{\hat i}(b)}{d\tau}.
 \eea
Thus, the solutions \cite{Crispino:2016pnv} are finally written as
 \bea
 \eta^{\hat 1}(r)&=&b\sqrt{\frac{2b}{m}}\frac{d\eta^{\hat 1}(b)}{d\tau}\left(\frac{b}{r}-1\right)^{1/2}
                +\eta^{\hat 1}(b)
                \left[\frac{3}{2}-\frac{r}{2b}+\frac{3}{4}\left(\frac{b}{r}-1\right)^{1/2}
                \cos^{-1}\left(\frac{2r}{b}-1\right)\right],\\
 \eta^{\hat i}(r)&=& r\left[\frac{\eta^{\hat i}(b)}{b}
               +\sqrt{\frac{2b}{m}}\frac{d\eta^{\hat i}(b)}{d\tau}\left(\frac{b}{r}-1\right)^{1/2}\right].
 \eea
These solutions of $\eta^{\hat 1}(r)$ and $\eta^{\hat i}(r)$ show
how the initial radial and angular separations of two nearby
geodesics are changed while falling to the Schwarzschild black
hole in massless gravity. It is appropriate to comment that
$\eta^{\hat 1}(b)$ and $\eta^{\hat i}(b)$ are initial separation
distances between two nearby geodesics at $r=b$ to the radial and
angular directions, respectively.  Moreover, $\frac{d\eta^{\hat
1}(b)}{d\tau}$ and $\frac{d\eta^{\hat i}(b)}{d\tau}$ are initial
velocities at $r=b$ to the radial and angular directions,
respectively, which mean either exploding when they are positive,
or imploding when they are negative. Here let us consider two
cases where one is $\eta^{\hat\alpha}(b)\neq 0$,
$\frac{d\eta^{\hat \alpha}(b)}{d\tau}=0$ $(\alpha=1,~2,~3)$, and
the other is $\eta^{\hat\alpha}(b)\neq 0$,
$\frac{d\eta^{\hat\alpha}(b)}{d\tau}\neq 0$. In Fig.
\ref{fig2}(a), (c) correspond to the former case, and (b), (d) to
the latter case. Note that when
$\frac{d\eta^{\hat\alpha}(b)}{d\tau}=0$, $\eta^{\hat\alpha}(r)$
become
 \bea\label{spaghetti}
 \eta^{\hat 1}(r)&=& \eta^{\hat 1}(b)
                \left[\frac{3}{2}-\frac{r}{2b}+\frac{3}{4}\left(\frac{b}{r}-1\right)^{1/2}
                \cos^{-1}\left(\frac{2r}{b}-1\right)\right],\\
 \eta^{\hat i}(r)&=& \frac{\eta^{\hat i}(b)}{b}r,
 \eea
respectively, which show that the separation distance goes to
infinity for the radial component due to the last term in
(\ref{spaghetti}) and goes to zero for the angular one as the body
is falling to the black hole, a process known well as
spaghettification.

On the other hand, when $\frac{d\eta^{\hat\alpha}(b)}{d\tau}\neq
0$, the radial components show the similar behaviors as in Fig.
\ref{fig2}(b), however the separation distances of the angular
components start to increase, reach a peak, then decrease to zero,
as the body falls to the black hole, which is shown in Fig.
\ref{fig2}(d). It is interesting to note that the separation
distance of the angular component is smaller as the mass of the
black hole is larger. Note also that by varying the initial
distances of $\eta^{\hat\alpha}(b)$ at $r=b$, the angular
component is more deformed as $\eta^{\hat i}(b)$ is larger.
\begin{figure*}[t!]
   \centering
   \includegraphics{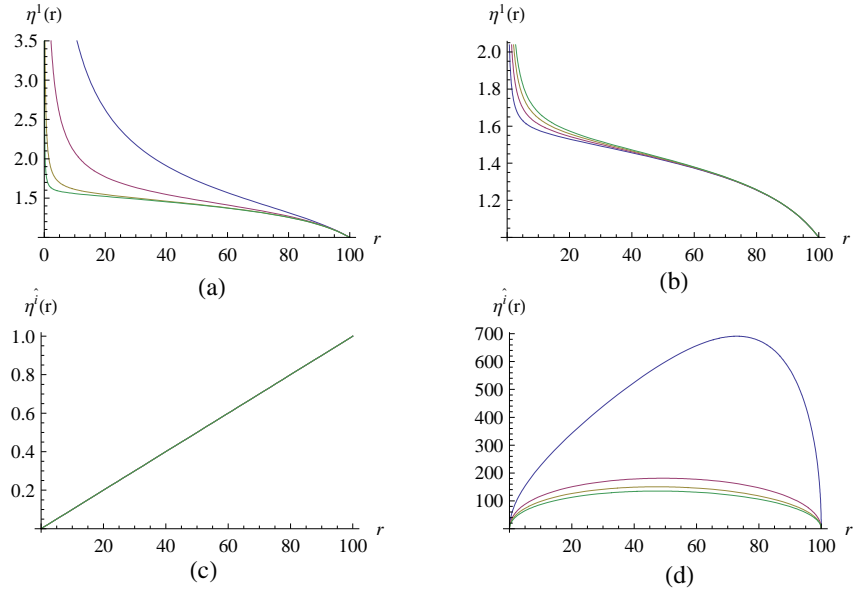}
\caption{For the Schwarzschild black hole in holographic massive
gravity, radial solutions of the geodesic deviation equation for
$\eta^{\hat 1}(b)=1$, $\frac{d\eta^{\hat 1}(b)}{d\tau}=0$: (a)
with $m=10$ by varying $R=0.001,~0.01,~0.1,~0.5$ from top to
bottom, (b) with $R=0.1$ by varying $m=5,~10,~15,~20$ from bottom
to top, and angular solution of the geodesic deviation equation,
(c) for $\eta^{\hat i}(b)=1$, $\frac{d\eta^{\hat i}(b)}{d\tau}$=0,
$m=10$ and  $R=0.001,~0.01,~0.1,~0.5$, and (d) for $\eta^{\hat
i}(b)=1$, $\frac{d\eta^{\hat i}(b)}{d\tau}$=1, and $R=0.001$ by
varying $m=5,~10,~15,~20$ from top to bottom.}
 \label{fig3}
\end{figure*}

Now, inspired by the massless case, let us solve the geodesic
deviation equations (\ref{diffm1}) and (\ref{diffm2}) by noting
that the constant terms of ${\cal C}$ are cancelled out and thus
expanding the integrand to the power of $R$ up to the $R^4$ terms
in order to see the massive graviton effect. Then, one can
integrate it out term by term as follows
 \bea
 \eta^{\hat 1}(r)&=&d_1\sqrt{\frac{2m}{b}}\left(1+\frac{bRr}{m}\right)^{1/2}
                 \left(\frac{b}{r}-1\right)^{1/2} \nonumber\\
              &+& d_2\frac{b}{2m}\left(1+\frac{bRr}{m}\right)^{1/2}
                \left[2bg_1(r)+\frac{3b}{2}g_2(r)\left(\frac{b}{r}-1\right)^{1/2}
                \cos^{-1}\left(\frac{2r}{b}-1\right)\right],\\
 \eta^{\hat i}(r)&=& d_3 r
               -d_4\sqrt{\frac{2b}{m}}\frac{r}{b}\left[g_3(r)\left(\frac{b}{r}-1\right)^{1/2}
               -g_4(r)\cos^{-1}\left(\frac{2r}{b}-1\right)\right],
 \eea
where
 \bea
 g_1(r)&=&
 \frac{3}{2}-\frac{r}{2b}+\left(\frac{3r^2}{8b}+\frac{15r}{16}-\frac{45b}{16}\right)\left(\frac{bR}{m}\right)
 -\left(\frac{5r^3}{16b}+\frac{35r^2}{64}+\frac{175br}{128}-\frac{525b^2}{128}\right)\left(\frac{bR}{m}\right)^2\nonumber\\
 &&+\left(\frac{35r^4}{128b}+\frac{105r^3}{256}+\frac{735br^2}{1024}+\frac{3675b^2r}{2048}-\frac{11025b^3}{2048}\right)\left(\frac{bR}{m}\right)^3\nonumber\\
 &&-\left(\frac{63r^5}{256b}+\frac{693r^4}{2048}+\frac{2079br^3}{4096}+\frac{14553b^2r^2}{16384}+\frac{72765b^3r}{32768}-\frac{218295b^4}{32768}\right)\left(\frac{bR}{m}\right)^4
 +{\cal O}(R^5),\nonumber\\
 g_2(r)&=&1-\frac{15b}{8}\left(\frac{bR}{m}\right)+\frac{175b^2}{64}\left(\frac{bR}{m}\right)^2
         -\frac{3675b^3}{1024}\left(\frac{bR}{m}\right)^3+\frac{72765b^4}{16384}\left(\frac{bR}{m}\right)^4+{\cal O}(R^5),\nonumber\\
 g_3(r)&=& 1+\frac{3br}{16}\left(\frac{bR}{m}\right)^2
         -\left(\frac{5br^2}{64}+\frac{15b^2r}{128}\right)\left(\frac{bR}{m}\right)^3
         +\left(\frac{35br^3}{768}+\frac{175b^2r^2}{3072}+\frac{175b^3r}{2048}\right)\left(\frac{bR}{m}\right)^4+{\cal O}(R^5),\nonumber\\
 g_4(r)&=&\frac{b}{4}\left(\frac{bR}{m}\right)-\frac{3b^2}{32}\left(\frac{bR}{m}\right)^2
         +\frac{15b^3}{256}\left(\frac{bR}{m}\right)^3-\frac{175b^4}{4096}\left(\frac{bR}{m}\right)^4+{\cal O}(R^5).
 \eea
Here, the constants of integration are given by
 \bea
 d_1&=&\frac{b^2}{m}\left(1+\frac{b^2R}{m}\right)^{-1}\frac{d\eta^{\hat 1}(b)}{d\tau},
        ~~d_2=\frac{m}{b^2}\left(1+\frac{b^2R}{m}\right)^{-1/2}g^{-1}_1(b)\eta^{\hat 1}(b),\nonumber\\
 d_3&=&\frac{1}{b}\eta^{\hat i}(b),
     ~~~~~~~~~~~~~~~~~~~~~~~~~d_4=-b\frac{d\eta^{\hat i}(b)}{d\tau}.
 \eea
Thus, the solutions are finally written as
 \bea
 \eta^{\hat 1}(r)&=&b\sqrt{\frac{2b}{m}}\left(1+\frac{b^2R}{m}\right)^{-1}
                 \frac{d\eta^{\hat 1}(b)}{d\tau}\left(1+\frac{bRr}{m}\right)^{1/2}
                 \left(\frac{b}{r}-1\right)^{1/2} \nonumber\\
              &+& \left(1+\frac{b^2R}{m}\right)^{-1/2}g^{-1}_1(b)\eta^{\hat 1}(b)\left(1+\frac{bRr}{m}\right)^{1/2}
                \left[g_1(r)+\frac{3}{4}g_2(r)\left(\frac{b}{r}-1\right)^{1/2}
                \cos^{-1}\left(\frac{2r}{b}-1\right)\right],\\
 \eta^{\hat i}(r)&=& r\left[\frac{\eta^{\hat i}(b)}{b}
               +\sqrt{\frac{2b}{m}}\frac{d\eta^{\hat
               i}(b)}{d\tau}\left(g_3(r)\left(\frac{b}{r}-1\right)^{1/2}
               -g_4(r)\cos^{-1}\left(\frac{2r}{b}-1\right)\right)\right].
 \eea

In Fig. \ref{fig3}, we have drawn the radial and angular
components of the separation vectors by varying $R$ and the mass
of the black holes, respectively. Here, we have found that the
massive gravitons keep tightening two nearby geodesics either as
the graviton's mass $\tilde{m}$ embodied in $R$ is bigger in Fig.
\ref{fig3}(a) or as the black hole's mass $m$ is smaller for a
given $R$ in Fig. \ref{fig3}(b). It is also interesting to see
that when $\frac{d\eta^{\hat i}(b)}{d\tau}=0$ the angular
component is linearly shrink to be zero as the body approaches the
singularity by
 \bea
  \eta^{\hat i}(r)&=& \frac{\eta^{\hat i}(b)}{b}r,
 \eea
regardless of the black hole's mass as shown in Fig.
\ref{fig3}(c). Moreover, when $\frac{d\eta^{\hat i}(b)}{d\tau}\neq
0$ with a fixed $R$, two nearby geodesics are more easily
separated as the mass of the black hole is smaller, then reach a
peak, decrease to zero, as the body falls to the black hole, as
seen in Fig. \ref{fig3}(d).
\begin{figure*}[t!]
 \centering
 \includegraphics{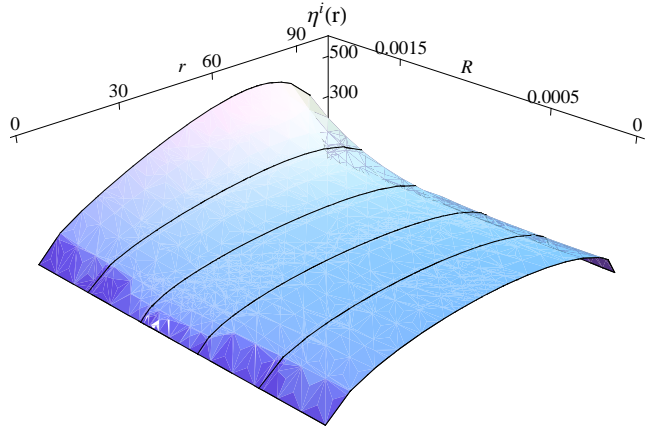}
  \caption{For the Schwarzschild black hole in holographic massive gravity,
angular solution of the geodesic deviation equation by varying $R$
with $\eta^{\hat 1}(b)=1$, $\frac{d\eta^{\hat 1}(b)}{d\tau}=1$ and
$m=10$.}
 \label{fig4}
\end{figure*}
In Fig. \ref{fig4}, we have also drawn the angular solution of the
geodesic deviation equation by varying the massive graviton
parameter $R$ continuously with fixed mass $m$ of the black hole.
This shows that as $R$ is increased, the distance between the
nearby geodesics in the angular direction is increased near the
departure point of $r=b$.

\section{Discussion}

In summary, we have studied the geodesics and tidal effects
produced in the spacetime of the Schwarzschild black hole in
Vegh's holographic massive gravity, which has two additional mass
parameters of $R$ and ${\cal C}$ due to the existence of massive
gravitons, comparing with the known results in massless gravity.
As a result, we have newly found that massive gravitons affect the
angular component of the tidal force giving an additional term
proportional to only $R/r$ coming from $f'(r)$, while the radial
component of the tidal force in holographic massive gravity is the
same with the one in massless gravity, which is derived from
$f''(r)$.

In order to see its implication, we have further investigated the
solutions of the geodesic deviation equations, which show the
behavior of two nearby geodesics freely falling into the
Schwarzschild black hole. As for the Schwarzschild black hole in
massless gravity, with boundary conditions of $\eta^{\hat
\alpha}(b)\neq 0$ and $\frac{d\eta^{\hat \alpha}(b)}{d\tau}=0$,
the radial component gets infinitely stretched and the angular
component is compressed to zero by the tidal force as shown in
Figs. \ref{fig2}(a) and \ref{fig2}(c). However, with exploding
boundary conditions of $\eta^{\hat \alpha}(b)\neq 0$ and
$\frac{d\eta^{\hat \alpha}(b)}{d\tau}\neq 0$, the separation
distances of the angular component start to get stretched, reach a
peak and then get compressed as shown in Fig. \ref{fig2}(d), while
the radial components behave similarly to the case of $\eta^{\hat
\alpha}(b)\neq 0$ and $\frac{d\eta^{\hat \alpha}(b)}{d\tau}=0$ as
in Fig. \ref{fig2}(b).

On the other hand, as for the Schwarzschild black hole in
holographic massive gravity, with boundary conditions of
$\eta^{\hat \alpha}(b)\neq 0$ and $\frac{d\eta^{\hat
\alpha}(b)}{d\tau}=0$, the radial components keep tightening and
after passing the event horizons get abruptly infinitely stretched
due to the massive gravitons, as shown in Fig. \ref{fig3}(a),
while the angular components are unaffectedly compressed as in
Fig. \ref{fig3}(c). With exploding boundary conditions of
$\eta^{\hat \alpha}(b)\neq 0$ and $\frac{d\eta^{\hat
\alpha}(b)}{d\tau}\neq 0$, the radial components also show the
abrupt stretch after passing the event horizons, while the angular
components are more deformed as the mass of a black hole is
smaller as shown in Figs. \ref{fig3}(b) and \ref{fig3}(d). As a
result, in the Schwarzschild black hole in holographic massive
gravity, we have newly shown that the massive gravitons keep
tightening radial components of two nearby geodesics as $R$ is
bigger. Moreover, the massive gravitons make angular components of
two nearby geodesics more deformed as the mass of the black holes
$m$ is smaller.

Finally, it seems appropriate to comment that all known massive
gravities may not be excluded by recent tests of GR in which
graviton mass bounds \cite{Will:1997bb} were continuously improved
in GW150914 \cite{Abbott:2016blz} and GW170104
\cite{Abbott:2017vtc} (see also
\cite{LIGOScientific:2019fpa,1826681}). Therefore, it would be
interesting to study more on various gravity models related with
tidal effects including the one in holographic massive gravity to
shed light on the nature of gravity.

\acknowledgments{S. T. H. was supported by Basic Science Research Program through
the National Research Foundation of Korea funded by the Ministry of Education, NRF-2019R1I1A1A01058449.
Y. W. K. was supported by the National Research Foundation of Korea grant funded by the
Korea government, NRF-2017R1A2B4011702.}


\end{document}